# A Better Definition of the Kilogram

by

Theodore P. Hill, School of Mathematics, Georgia Institute of Technology
Ronald F. Fox, School of Physics, Georgia Institute of Technology
and
Jack Miller, Lawrence Berkeley National Laboratory

## 1. Introduction

The current International System of Units (SI) definition of the kilogram officially recognized by the International Committee for Weights and Measures (CIPM - *Comite' International des Poids et Mesures*) is

(D1) A kilogram is the mass of the International Prototype Kilogram (*IPK*).

Even more precisely, it is the mass of that unique prototype artifact *IPK* "immediately after cleaning by a prescribed technique" [3, p 2251].

This definition (D1) is based on a unique 120-year old platinum-iridium cylinder, and even though there is no incontrovertible proof [3, p 2262], many experts believe that the mass of the *IPK* is changing in time [12, p 237]. For that, and numerous other reasons (e.g., see lay articles [1, 2], and scientific articles [3, 11, 14]), it is now widely accepted that the kilogram should be redefined in more modern scientific terms. The replacement of the definition (D1) by an intrinsic scientific (non-artifact) definition has thus been deemed a high priority.

Two recent articles in *Metrologia* by representatives of the International Bureau of Weights and Measures - *Bureau International des Poids et Mesures* (BIPM), and the U.S. National Institute of Standards and Technology (NIST), including the head of the Consultative Committee on Units (CCU) advisory committee to BIPM and the head of the Fundamental Constants Data Center at NIST, have proposed new definitions of the kilogram and other SI units [11, 12]. Since the perceived weakness of the current SI definitions of other units such as the ampere, mole, and candela "derives in large part from their dependence on the kilogram…the definition of the kilogram is thus central to the more general problem of improving the SI" [12, p 228]. Accordingly, the CCU and CIPM have

> *called for the widest possible publicity to be given to these ideas among the scientific and user communities so that their reactions and views can be taken* *into account in a timely way*…[and] *for a wide discussion to take place* [12, p 228-229].

This present paper is a response to that invitation to enter into the discussion about the BIPM/NIST proposal [12], and to describe the views and reactions of some scientists and users who are not metrologists. Our goal is to discuss a recent alternative redefinition of the kilogram [4], and to compare the respective definitions.

## 2. Proposed Definitions

In [11], the CCU/NIST authors describe two possible methods for redefinition of the kilogram: one based on fixing the value of Planck's constant $h$, and then using a watt-balance method to realize this definition; and the other fixing the Avogadro constant $N_A$ and then using a silicon-sphere (X-ray crystal diffraction) method to realize the definition.

In a subsequent article [12], the same authors settle on the fixed-$h$ method, which is based on two fundamental physics equations: $E = mc^2$ and $E = hf$, the first from special relativity theory and the second from quantum mechanics. Using both equations to solve for $m$ yields $m = hf/c^2$, and solving for $f$ yields $f = mc^2/h$; using the latter they propose three concrete definitions, namely,

(D2a) *The kilogram is the mass of a body whose equivalent energy is equal to that of a number of photons whose frequencies sum to exactly* $(299792458^2/66260693) \times 10^{41}$ *hertz.*

(D2b) *The kilogram is the mass of a body whose de Broglie–Compton frequency is equal to exactly* $299792458^2/(6.6260693 \times 10^{-34})$ *hertz.*

(D2c) *The kilogram, unit of mass, is such that the Planck constant is exactly* $6.6260693 \times 10^{-34}$ *joule second.*

Definitions (D2a) and (D2b) are "explicit-unit definitions", and (D2c) is an "explicit-constant definition"; see [12, p 233-234] for details and explanations. The authors also propose redefinitions of other SI units, although those necessitate the introduction of a new *additional fundamental constant kappa* (a correction factor) that is very close to zero, and which one author says is changing in time [NIST emails 2/18/07, 5/4/07], and another author says is not changing in time [NIST email 10/23/07]. Here we will not discuss the other aspects of the proposal in [12], but will restrict attention to the redefinition of the kilogram.

Our own proposed redefinition of the kilogram is a simple, concrete version of the method of fixing the Avogadro constant $N_A$, namely

(D3) A kilogram is the mass of $2250 \times 28148963^3$ atoms of carbon-12 at rest and in their ground state.

Other choices for the exact numerical value of the number of carbon-12 atoms in a kilogram are also possible, such as one which specifies an exact number of atoms in a gram (see Section 4 below); definition (D3) was chosen to make the mole and the Avogadro constant particularly simple, and because it is very close to the current recommended values (see Section 5 below).

As explained clearly in [7], from the equation relating Planck's constant, the fine structure constant, the Rydberg constant, the speed of light and the electron mass, it is possible to fix exactly any two of the three fundamental constants Avogadro's number, Planck's constant, and the carbon-12 molar mass. Since Planck's constant is considered one of the most fundamental of all physical constants, in addition to (D3) we also propose simultaneously fixing the exact value of Planck's constant. (For arguments supporting fixing the carbon-12 molar mass, see [7].)Thus, in analogy to (D2c)(see [12, Table 1, row 3]), we also propose fixing $h$ so that

(4). *The kilogram is such that the Planck constant is exactly* $6.6260693 \times 10^{-34}$ *joule second.*

It is very important to note here that, unlike (D2c) above, (4) is *not a definition* of the kilogram, but rather is a consequence of the definition (D3) and the new proposed fixed value (zero uncertainty) for $h$ given in (4).

## 3. Practical realizations

Once a new definition of the kilogram such as (D2a-b-c) or (D3) is adopted, no man-made object (including the *IPK*) will ever have mass exactly one kilogram, except by pure chance and then for only a few nanoseconds. However, there still will be a need for practical realizations of the kilogram similar to the various national copies of the *IPK*. Initially, the existing prototype copies will serve that role, but as time goes on, more accurate copies will certainly be needed, and perhaps more countries will want copies. In this section, we will review some of the issues in actually realizing prototype kilograms under the new definition, in other words, constructing or calibrating scales that are highly accurate.

As the CCU/NIST team says, the long-term goal is to design

> *a comparatively easy-to-use apparatus that can enable the experimental realization of the new definition of the kilogram with the appropriate uncertainty at any place at any time by anyone* [12, p 238].

That is exactly what happened in 1983 when the *meter* was redefined as the distance light travels in exactly 1/299,792,458 seconds, thereby eliminating the need for the official artifact platinum-iridium meter stick forever. Since that redefinition, any student with a stopwatch, laser pointer, strobe light and rotating mirror can construct a reasonably accurate “meter stick” independent of any other prototype. At the other extreme, each national laboratory that so desires can now construct a meter stick to whatever accuracy it is willing to expend the resources for.

**Definitions (D2a-b-c).**

Although definitions (D2a-b-c) are attractive from the standpoint of theoretical physics, unfortunately the watt-balance experiment needed to make practical realizations of a kilogram using themis indirect, and does not produce an actual object of a given mass, i.e. a kilogram artifact.

Moreover, the watt-balance method requires substantial resources, hence their rare status. The one at NIST is two stories high, cost over US$1.5 million to set up, and requires a team of between three and five expert scientists working on the project at any one time, as well as considerable use of expensive liquid helium for the two superconducting magnets. In addition, according to the head of the NIST watt-balance unit, the problem with watt-balances is they can be finicky – distant earthquakes, motors from nearby offices and tides have shaken up the measurements, and there are 20 potential sources of error, including buoyancy of air, electrical current leaks, and changes in local gravity . The National Physics Laboratory in the U.K., the site of the first watt-balance, recently ceased funding of its own watt-balance experiments, dismantled the device, and shipped it to Canada. Also, there are not many scientists with the expertise to build and run a watt-balance, and not many countries are willing to afford such expenses year after year [1],[2].

The problem of coping with the various sources of error in the watt-balance experiment is compounded dramatically by the magnitudes specified in the definitions. Definitions (D2a-b) entail a de Broglie-Compton frequency in *hertz* (cycles per second) of more than $10^{50}$ *hz*. With the enormity of this frequency, it is not easy to understand how this will be accurately measured in practice, since the cesium (cesium-133) clock frequency to determine the second is only *9,192,770 hz*, or about $9 \times 10^{9}$*hz*, so the defining frequency in (D2a-b) is 41 orders of magnitude greater than that of the number of vibrations defining the second. Moreover, the proposed exact constant in (D2a) and (D2b) is not only larger than $10^{50}$, but also is a non-terminating (infinite) decimal, which is still only an approximation even when rounded to a billion digits. In [10, p 2] the authors maintain that “It would be inconvenient to quote concentrations…with numbers of the order $10^{23}$…”, but definitions (D2a) and (D2b) even require estimates of frequencies 27 orders of magnitude greater than that. Similarly, the explicit-constant definition (D2c) yields a mass (via the above equation $m = hf/c^2$) many orders removed from a kilogram ($\cong 4 \times 10^{-41}$*kg*).

At the grassroots scientific level, it is difficult for students and even university professors to perform simple laboratory experiments based on definitions (D2a-b-c) to construct a rough approximation of a kilogram mass, as the redefinition of the meter permitted a standard school laboratory experiment to construct a rough meter stick.

**Definition (D3).**

Adopting a fixed value for Planck's constant (e.g. using the center of the currently recommended NIST value as in (4)), one *indirect* method of practical realization of (D3) is to use a watt-balance experiment and the equivalence that (4) provides, albeit with the reservations above.

The most *direct* practical realization of (D3) is currently available through several laboratories in the Avogadro Project, where silvery soft-ball-sized artifacts of single-crystal silicon spheres of high purity and nearly spherical shape yield an estimate of the number of silicon-28 atoms in a given macroscopic mass. The first such results were published in 2004, and currently only two such spheres exist, each costing about $3.2 million and manufactured and maintained by master opticians [1, 2]. The objective of this method is simply to estimate the number of atoms in the sphere, using estimates of the imperfect purity of the silicon isotopes, the average volume of a silicon-28 atom, the radius of the rough sphere, and the mass of the sphere. Thus the sphere itself is a sophisticated practical realization of an atom-counting definition such as in (D3).

In contrast to definitions (D2a-b-c), definition (D3) also allows a simple rough prototype of a kilogram mass to be constructed in a school laboratory, or even at home: a block of nearly pure carbon, cut so that it is roughly 8.11 cm (or as close to 368,855,762 carbon-12 atoms as possible) on a side, will be approximately one kilogram. Of course, the exact dimensions depend on the form of carbon used – graphite, say, or diamond – and on its crystal lattice structure (cf. [4]). At this point in time, it is not yet possible to obtain exact counts of individual atoms, even when they are in a crystal lattice, but that is merely a question of time.

## 4. Uncertainties

One of the primary criteria in selecting a redefinition of the kilogram (and other SI units) is that the new definition be consistent with other current SI definitions and with current accepted values of the fundamental constants; as [12, p 228] states:

> *if the definition is to replace an earlier definition of the same unit, it should be chosen to preserve continuity, so that the new definition should be consistent with the previous definition within the uncertainty that the previous definition could be realized.*

The term *uncertainty* in the SI setting has a very particular meaning – (relative) standard uncertainty is the estimated (relative) standard deviation of the variable – and in practice any value outside the published uncertainty of a constant has been rejected out of hand. For example, some of the previous proposed redefinitions of the kilogram have been criticized because they do not fit within exactly one "uncertainty". The kilogram entry on Wikipedia, for instance, states "the Avogadro constant has a relative standard uncertainty of 50 parts per billion and …*Unfortunately*, none of the three integer values within the range possess the property of their cubes being divisible by twelve" [15, emphasis added]. Thus to examine whether proposed redefinitions of the kilogram and other SI units meet the CCU goal of consistency with current values, it is important to examine and understand exactly what published recommended values of "uncertainty" mean.

The standard uncertainty of the NIST recommended value of a fundamental physical constant is one of two types: Type A - an evaluation based on a *statistical analysis* of data; and Type B - *all others*, including previous measurement data, manufacturer's specifications, and experience with, or general knowledge of, the behavior and property of relevant materials and instruments. When the type of uncertainty (A or B) for the CODATA recommended value of a fundamental constant is known, that type is supposedly specified. Almost none are.

From a statistical standpoint, knowing only the value of the standard deviation of a variable gives very little useful information without knowledge of the underlying distribution. For some distributions, 90% of the values lie *outside* one standard deviation. The statement "within one standard deviation" is commonly interpreted as including 67% of the values, the value for a Gaussian variable. However, the underlying data distributions for the fundamental physical constants are usually not known or are not reported, and are never true Gaussian distributions (which always include negative values that are usually, but not always, negligible). In calculation of standard deviations, NIST also uses a technique that linearizes the relevant equations and estimates the effects of covariances between variables, and these, too, introduce additional errors. Thus, as members of the CODATA Task Force on Fundamental Constants conceded, it is "difficult enough to evaluate uncertainties in a meaningful way" [13, p 53].

Moreover, official recommended values for uncertainty are subject to *ad hoc* change by the committee in charge. For example, when data analysis uncovered two major discrepancies with the 2006 input data, the CODATA Task Group decided to weight the *a priori* assigned uncertainties of all five data by the multiplicative factor 2.325 which "reduced the discrepancies to a level comfortably below two standard uncertainties" [12], [13]. As the CODATA Task Force concluded, "the 2006 CODATA set of recommended values…does not rest on as solid foundation as one might wish." [13, p 54]. The fundamental problem of how to combine data from independent experiments, especially experiments of different types such as watt-balance and crystal X-ray diffraction, is indeed a difficult one, but recent advances in statistical theory may well prove very useful (cf. [5], [6]).

Historically, the official uncertainties have not been improving (decreasing) steadily as one might expect - the official NIST recommended relative uncertainty for the Avogadro constant was 360 parts per billion (ppb) in 1986, 47 ppb in 1998, 100 ppb in 2002, and 30 ppb in 2006 – and there is no reason to expect smaller uncertainty in 2010.

The lesson here is that in selecting new definitions for the kilogram and other SI units, strict adherence to values within 1 or even 2 uncertainties of the current recommended values is not justifiable. A more reasonable benchmark is laid out by the CCU/NIST team in [11, p 75]: "we believe that even if it were to be eventually discovered that the value of $h$ or $N_A$ chosen to redefine the kilogram were such that [the relative deviation in mass of the new kilogram from the mass of *IPK*] $\approx 10^{-6}$… the consequences could be better dealt with through a redefinition now."

In terms of the proposed redefinition of the kilogram, this means that the range of acceptable values for the mass of a kilogram in terms of exact numbers of atoms (D3) is many times more than one relative standard uncertainty, and could easily be fixed at a perfect cube divisible by 12, namely, $84446892^3$, so that one gram is exactly $144 \times 7037241^3$ Carbon-12 atoms. As mentioned above, the selection of the value in (D3) was based on being extremely close to current recommended values, and choosing anything reasonably close to that value or to the recommended value at the time of redefinition of the kilogram will suffice.

## 5. Educational Aspects

One of the most crucial considerations of any redefinition of the kilogram and other SI units is the legacy we leave to the next generation of scholars. As [12, p228] declares

> *since it is important that the basis of our measurement system be taught in schools and universities, it is preferable, as far as modern science permits, that the definitions of base units be comprehensible to students in all disciplines*.

**Definitions (D2a-b-c).**

To understand these three proposed redefinitions of the kilogram requires knowledge of physics at the advanced university or even graduate level, including special relativity and quantum mechanics. Since we found this indirect approach to redefine the kilogram difficult to understand, we asked the authors of [12] what their introductory-level textbook definition of a kilogram would be, including all the necessary pre-definitions such as de Broglie, photon frequency, etc. The response, from a member of the CODATA Task Group on Fundamental Constants, was "I am not in the business of writing introductory textbooks -- I will leave that to others. All I will say is that we do not believe any of the proposed new definitions are any more complex than the current definitions of some of the SI base units" [NIST email 5/4/07]. Our similar query to the NIST watt-balance team was answered with a long paragraph, but no clean concise definition [NIST email 4/23/08].

We feel that if any of the definitions (D2a-b-c) is eventually adopted, that definition, as [12] said, should be made "comprehensible to students in all disciplines", as far as modern science permits.

**Definition (D3).**

Modern science *does* permit a simple redefinition of the kilogram that is easily comprehensible to students in all disciplines, and (D3) is one such example. Students need only have an idea what an atom of carbon-12 is.

Moreover, and equally important, (D3) also allows clean and concise definitions of the Avogadro constant $N_A$ and the mole, namely,

(D4). *Avogadro's constant*, the number of atoms in 12 grams of carbon-12, is

$$N_A = 84446889^3 = 602{,}214{,}162{,}464{,}240{,}116{,}093{,}369.$$

and

(D5). The *mole* is the amount of substance that contains exactly $84446889^3$ specified elementary entities, which may be atoms, molecules, ions, electrons, other particles or specified groups of such particles.

The main reason for defining the Avogadro constant and the mole in terms of numerical cubes is that unlike the five other SI base quantities (length, time, electric current, temperature, and luminous intensity), both mass and mole are inherently 3-dimensional: if a region contains positive mass or mole of a substance, then it must contain a 3-dimensional geometrical cube. This stems easily from the standard mathematical fact that a region has positive volume if, and only if, it contains a geometrical cube. Moreover, the base (side length) of the proposed defining numerical cubes in $N_A$ and the mole are eight significant digits, and thus exactly consistent with current measurement goals and standards. Also, a cube is much easier to visualize than a fourth or higher power, and, although easy to visualize, no geometrical square can contain any positive mass or mole. Other reasons for choice of a perfect cube, in addition to its elegance and simplicity, are outlined in [4].

A simple, clean definition of the mole is especially important in chemistry. The Chair of the Committee on Nomenclature, Terminology and Symbols of the American Chemical Society, wrote that by simply fixing an integral value for Avogadro's number in this manner, "much of what seems to confuse many students about the mole in introductory courses will be dampened" [8]. And, as confirmed in an earlier article by the same proposers of definitions (D2a-b-c), an atom-counting definition of the kilogram that fixes $N_A$ (as (D3) and (D4) do) "is simple, conceptually, enabling it to be widely

understood…[and] allows the mole to be redefined in a simpler and more understandable way" [11, p 77].

One of the official recommendations of the International Committee for Weights and Measures was to "further encourage National Metrology Institutes to pursue national funding to support continued relevant research" [12, p 245], and in making a decision to eliminate an easily-understood atom-counting definition in favor of a watt-balance definition (D2a-b-c), even the appearance of ulterior funding considerations should be carefully avoided.

## 6. Conclusions

In our opinion, the proposed new definition of the kilogram (D3) is: (i) more experimentally neutral than definitions (D2a-b-c), which heavily favor the sensitive watt-balance experiments; (ii) much easier for school and university science laboratories to use to make rough realizations than (D2a-b-c); and, of utmost importance to future generations who will use the SI, (iii) vastly easier to comprehend and visualize than (D2a-b-c).

Although the head of the Consultative Committee on Units to the BIPM has insisted that "the time [for redefinition of the kilogram] is not only right, but urgent", has submitted his own proposal for redefinition (D2a-b-c), and has said "I am convinced we've got it right" [1, p 49], we hope that his committee remains open to the alternative proposal (D3).

The head of the Mass Section at BIPM, who is cautious about the 2011 deadline, said "It's not yet urgent. …people have been living with this for years" [1, p 49] and it is this view that we share. But if a decision is to be made, we strongly prefer the elegant redefinition (D3).

**Acknowledgements.** The authors are grateful to Peter Becker, Richard Davis, Paul Karol, Peter Mohr, Richard Steiner, and Barry Taylor for helpful communications; the NIST emails cited in the text are available on request.

## References

**[1]** Bowers, M.(2009). Why the world is losing weight. *The Caravan* September 1-15, 42-49.

**[2]** Chong, J-R. (2008) Kilogram issue weighs on world, *Los Angeles Times* April 17, 2008, A-1 and A-20.

**[3]** Davis, R. (2005) Possible new definitions of the kilogram, *Phil. Trans. R. Soc. A* 363, 2249- 2264.

**[4]** Fox, R.F., Hill, T. (2007). An Exact Value for Avogadro's Number, *American Scientist* 95, 104-107 (March-April 2007).

**[5]** Hill, T. (2010) Conflations of probability distributions, accepted for publication in *Transactions of the American Mathematical Society*, http://arxiv.org/abs/0808.1808

**[6]** Hill, T., Miller, J., and Fox, R.F. (2010) How to combine independent data sets for the same quantity, http://arxiv.org/abs/1005.4978

**[7]** Jeannin, Y. (2010). What is a mole?:Old concepts and new continued. *Chemistry International* Vol 32.

**[8]** Karol, P. (2007), Letter to the Editors, *American Scientist* 95, 291.

**[9]** Lorimer, J. (2010) What is a mole?:Old concepts and new. *Chemistry International* Vol 32.

**[10]** Mills, I. and Milton, M. (2009) Amount of substance and the mole, *Chemistry International* Vol 31.

**[11]** Mills, I., Mohr, P., Quinn, T., Taylor, B., and Williams, E. (2005). Redefinition of the kilogram: A decision whose time has come. *Metrologia* 42:71-80.

**[12]** Mills, I., Mohr, P., Quinn, T., Taylor, B., and Williams, E. (2006). Redefinition of the kilogram, ampere, Kelvin, and mole: a proposed approach to implementing CIPM recommendation 1 (CI-2005), *Metrologia* 43, 227-246.

**[13]** Mohr, P., Taylor, B. and Newell, D. (2007) The fundamental physical constants, *Physics Today*, 52-55.

**[14]** Robinson, I. 2006. Weighty matters. *Scientific American* 295(b):102-109.

**[15]** Wikipedia entry on “Kilogram”, http://en.wikipedia.org/wiki/Kilogram accessed May 1, 2010.